\documentclass[12pt]{article}

\usepackage{sbc-template}

\usepackage{graphicx,url}

\usepackage{authblk}

\usepackage[latin1]{inputenc}  

\newcommand{\aspas}[1]{{``#1''}}
\newcommand{\mcode}[1]{$\tt #1$}

\usepackage{listings}
\usepackage{color}

\definecolor{backcolor}{gray}{0.9}

\definecolor{darkgray}{rgb}{.4,.4,.4}
\definecolor{purple}{rgb}{0.65, 0.12, 0.82}

\lstdefinelanguage{JavaScript}{
	keywords={typeof, new, true, false, catch, function, return, null, catch, switch, var, if, in, while, do, else, case, break, constructor, class},
	keywordstyle=\color{blue}\bfseries,
	ndkeywords={class, export, boolean, throw, implements, import, this},
	ndkeywordstyle=\color{darkgray}\bfseries,
	identifierstyle=\color{black},
	sensitive=false,
	comment=[l]{//},
	morecomment=[s]{/*}{*/},
	commentstyle=\color{purple}\ttfamily,
	stringstyle=\color{red}\ttfamily,
	morestring=[b]',
	morestring=[b]"
}

\title{JSClassFinder: A Tool to Detect Class-like \\ Structures in JavaScript}

\author{Leonardo Humberto Silva\inst{1}, Daniel Hovadick\inst{2}, Marco Tulio Valente\inst{2}, 
	Alexandre~Bergel\inst{3},Nicolas~Anquetil\inst{4}, Anne Etien\inst{4}}

\address{Department of Computing -- Federal Institute of Northern Minas Gerais (IFNMG)\\
  Salinas -- MG -- Brazil
\nextinstitute
  Department of Computer Science -- Federal University of Minas Gerais (UFMG)\\
  Belo Horizonte -- MG -- Brazil
\nextinstitute
  Department of Computer Science -- Pleiad Lab\\
  University of Chile -- Santiago -- Chile
\nextinstitute
  RMoD Project-Team -- INRIA Lille Nord Europe -- France
\email{leonardo.silva@ifnmg.edu.br, \{dfelix,mtov\}@dcc.ufmg.br, abergel@dcc.uchile.cl,  
	\{nicolas.anquetil,anne.etien\}@inria.fr}}

\begin{document} 

\maketitle

\begin{abstract}
With the increasing usage of JavaScript in web applications, there is a great demand to write JavaScript code that is reliable and maintainable. To achieve these goals, classes can be emulated in the current JavaScript standard version. In this paper, we propose a reengineering tool to identify such class-like structures and to create an object-oriented model based on JavaScript source code. The tool has a parser that loads the AST (Abstract Syntax Tree) of a JavaScript application to model its structure. It is also integrated with the Moose platform to provide powerful visualization, e.g., UML diagram and Distribution Maps, and well-known metric values for software analysis. We also provide some examples with real JavaScript applications to evaluate the tool.
\end{abstract}

\noindent Video: \url{http://youtu.be/FadYE_FDVM0}
     
\section{Introduction} \label{sec:introduction}

JavaScript is a loosely-typed dynamic language with first-class functions. It supports object-oriented, imperative, and functional programming styles. Behaviour reuse is performed by cloning existing objects that serve as prototypes~\cite{guha2010}. ECMAScript~\cite{ecmascript51} is a scripting language, standardized by ECMA International, that forms the base for the JavaScript implementation. ECMAScript 5 (ES5) is the version currently supported by most browsers. Version 6 of the standard is planned to be officially released around mid 2015\footnote{https://developer.mozilla.org/en-US/docs/Web/JavaScript, verified 05/18/2015}.

Due to the increasing usage of JavaScript in web applications, there is a great demand to write JavaScript code that is reliable and maintainable. In a recent empirical study, we found that many developers emulate object-oriented classes to implement parts of their systems~\cite{silva-saner2015}. However, to the best of our knowledge, none of the existing tools for software analysis of JavaScript systems identify object-oriented entities, such as classes, methods, and attributes. 

In this paper, we propose and describe the JSClassFinder reengineering tool that identifies class-like structures and creates object-oriented models based on JavaScript source code. Although ES5 has no specific syntax for class declaration, JSClassFinder is able to identify structures that emulate classes in a system. Prototype-based relationships among such structures are also identified to infer inheritance. The resulting models are integrated with Moose\footnote{http://www.moosetechnology.org/, verified 05/18/2015}, which is a platform for software and data analysis. The main features of the proposed tool are:

\begin{itemize}
	\item Identification of class-like entities to build an object-oriented model of a system.
	\item Integration with a complete platform for software analysis.
	\item Graphical visualization of the retrieved class-like structures, using UML class diagrams, distribution maps, and tree view layouts.
	\item Automatic computation of widely known source code metrics, such as number of classes (NOC), number of methods (NOM), number of attributes (NOA), and depth of inheritance tree (DIT).  
\end{itemize}

This tool paper is organized as follows. Section~\ref{sec:overview} describes JSClassFinder's architecture. Section~\ref{sec:examples} uses one toy example and two real applications to demonstrate the tool. Section~\ref{sec:limitations} describes exceptions that are not covered by the tool. Section~\ref{sec:related} presents related work and Section~\ref{sec:conclusions} concludes the paper.

\section{JSClassFinder in a Nutshell} \label{sec:overview}

The execution of JSClassFinder is divided into two stages: preprocessing and visualization. The preprocessing is responsible for analyzing the AST of the source code, identifying class-like entities, and creating an object-oriented model that represents the code. In the visualization stage, the user can interact with the tool to visualize the model and to inspect all metrics and visualization features that the software analysis platform provides. 

JSClassFinder is implemented in Pharo\footnote{http://pharo.org/, verified 05/18/2015}, which is a complete Smalltalk environment for developing and executing object-oriented code.
Pharo also offers strong live programming features such as immediate object manipulation, live update, and hot recompilation. The system requirements to execute JSClassFinder are: (i) the AST of a JavaScript source code in JSON format; (ii) A Pharo image with JSClassFinder. A ready-to-use Pharo image is available at the JSClassFinder website\footnote{http://aserg.labsoft.dcc.ufmg.br/jsclasses/, verified 05/18/2015}. Figure~\ref{fig:architecture} shows the architecture of the JSClassFinder, which includes the following modules:

\begin{figure}[ht]
	\small
	\centering
	\includegraphics[width=0.9\textwidth]{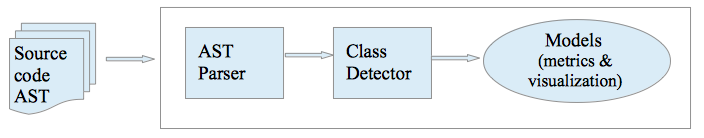}
	\caption{JSClassFinder's architecture}
	\label{fig:architecture}
\end{figure}

\noindent \textbf{AST Parser:} This module receives as an input the AST of a JavaScript application. Then it creates a JavaScript model as part of the preprocessing stage. Currently, we are using Esprima\footnote{http://esprima.org/, verified 05/18/2015}, a ECMAScript parser, to generate the AST in JSON format. 

\noindent \textbf{Class Detector:} This module is responsible for identifying class-like entities in the JavaScript model. It is the last step of the preprocessing stage, when an object-oriented model of an application is created and made available to the user.

\noindent \textbf{Models (metrics \& visualization):} This module provides visualizations for an user to interact with the tool and to \aspas{navigate} by the application's model. All information about classes, methods, attributes, and inheritance relationships is available. The main visualizations provided are: UML class diagram, distribution maps~\cite{ducasse-icsm2006}, and tree views. For the metrics, the tool  provides the total number of classes (NOC) and, for each class: number of methods (NOM), number of attributes (NOA), number of children (subclasses) and depth of inheritance tree (DIT).

\begin{figure}[ht]
	\centering
	\includegraphics[width=0.6\textwidth]{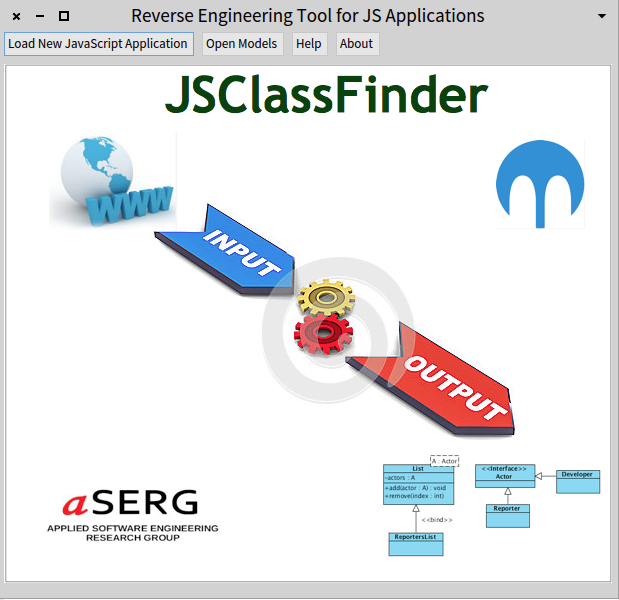}
	\caption{JSClassFinder initial user interface}
	\label{fig:interface}
\end{figure}

Figure~\ref{fig:interface} shows the main browser of JSClassFinder's user interface. In the top menu, the user can load a new JavaScript application or open one existing model, previously loaded. The only information required to load new applications are: (i) application's name and (ii) root directory where the JSON files with the target system's AST are located. After the preprocessing stage, the tool opens a new model inside a panel, where the user can navigate through class entities and select any graphical visualizations or metrics available. Section~\ref{sec:examples} presents some examples of what the user can do once they have a model created. 

\subsection{Strategy for class detection} \label{sec:strategy}

In this section, we describe the strategy used by JSClassFinder to represent the way classes are created in JavaScript and the way they acquire fields and methods. A detailed description is provided in a conference paper~\cite{silva-saner2015}.

\noindent{\em Definition \#1}: A class is a tuple $(C, \mathcal{A}, \mathcal{M})$, where $C$ is the class name, $\mathcal{A} = \{a_1, a_2, \ldots, a_p\}$ are the attributes defined by the class, and $\mathcal{M} = \{m_1, m_2, \ldots, m_q\}$ are the methods. Moreover, a class $(C, \mathcal{A}, \mathcal{M})$, defined in a program $P$, must respect the following conditions:

\begin{itemize}
	
	\item $P$ must have a function with name \mcode{C}.
	
	\item $P$ must include at least one expression of type \mcode{new} \mcode{C()} or \mcode{Object.create} \mcode{(C.prototype)}.
	
	\item For each $a \in \mathcal{A}$, the function \mcode{C} must include an assignment \mcode{this.a= Exp} or $P$ must include an assignment \mcode{C.prototype.a= Exp}. 
	
	\item For each $m \in \mathcal{M}$, function \mcode{C} must include an assignment \mcode{this.m= function} \mcode{\{ Exp \}} or $P$ must include an assignment \mcode{C.prototype.m= function} \mcode{\{ Exp \}}.\\[-.15cm]
\end{itemize}

\noindent{\em Definition \#2}: Assuming that $(C1,\mathcal{A}1,\mathcal{M}1)$ and $(C2,\mathcal{A}2,\mathcal{M}2)$ are classes in a program $P$, we define that $C2$ is a subclass of $C1$ if one of the following conditions holds:

\begin{itemize}
	\item $P$ includes an assignment \mcode{C2.prototype=} \mcode{new} \mcode{C1()}.
	\item $P$ includes an assignment \mcode{C2.prototype=} \mcode{Object.create(C1.prototype)}.
\end{itemize}


\section{Examples} \label{sec:examples}

This section shows examples of usage of JSClassFinder to analyze one toy example and two real JavaScript applications extracted from GitHub.

\subsection{Toy Example}  \label{sec:toy-example}

This example includes two simple classes, \mcode{Mammal} and \mcode{Cat}, to illustrate how classes can be emulated in JavaScript. Listing~\ref{lst_toy-example} presents the function that defines the class  \mcode{Mammal} (lines 1-3), which includes an attribute \mcode {name}. This class also has a method named \mcode{toString} (lines 4-6), represented by a function which is associated to the prototype of \mcode{Mammal}. Line 7 indicates that the class \mcode{Cat} (lines 9-11) inherits from the prototype of \mcode{Mammal}. The usage of variables \mcode{animal} and \mcode{myPet} demonstrate how the classes can be instantiated and used (lines 12-13). 

\begin{figure}[htbp]
	\begin{lstlisting}[caption=Class declaration and object instantiation, label=lst_toy-example]
function Mammal(name) { 
	this.name=name;
} 
Mammal.prototype.toString=function(){ 
	return '['+this.name+']';
} 
Cat.prototype = Object.create(Mammal.prototype);  // Inheritance
...       
function Cat(name) { 
	this.name='"meow" ' + name;
} 
var animal = new Mammal('Mr. Donalds');
var myPet = new Cat('Felix');
	\end{lstlisting}
\end{figure}

Listing~\ref{lst_toy-example} represents one way of defining classes and instantiating objects. There are some variations and customized implementations, as we can find in~\cite{flanagan2011, crockford2008, gama-wse2012}. JSClassFinder supports different types of class implementation, as reported in~\cite{silva-saner2015}. 

\subsection{Algorithms.js}

\begin{figure}[ht]
	\centering
	\includegraphics[width=1\textwidth]{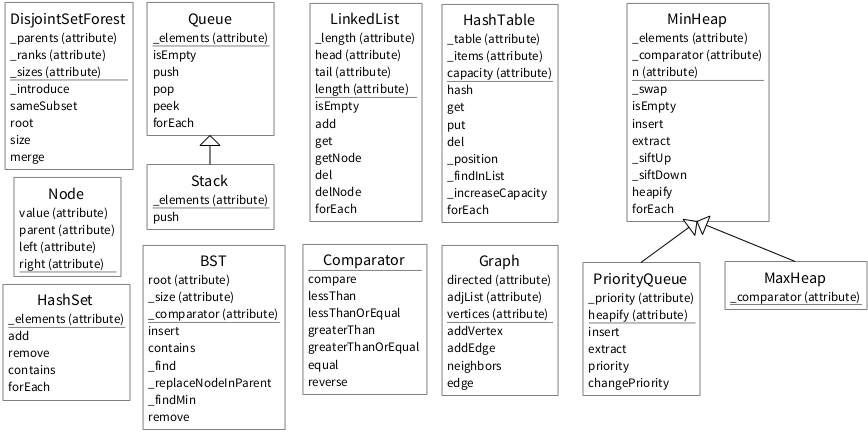}
	\caption{UML class diagram for Algorithms.js}
	\label{fig:algorithms}
\end{figure}

\mcode{Algorithms.js}\footnote{https://github.com/felipernb/algorithms.js/ verified 05/18/2015} is an open source project that offers traditional algorithms and data structures implemented in JavaScript. We analyzed version 0.8.1, which has 3,263 LOC. Figure~\ref{fig:algorithms} shows the class diagram of Algorithms.js, generated automatically by JSClassFinder, after the preprocessing stage. The classes represent the data structures that are supported by Algorithms.js, as we can check in the project's documentation webpage\footnote{http://algorithmsjs.org/ verified 05/18/2015}. The main algorithms, e.g., Dijkstra, EulerPath, Quicksort, that the application offers in its API, are implemented as JavaScript global functions, not classes.

\subsection{PDF.js}

\mcode{PDF.js}\footnote{https://github.com/mozilla/pdf.js verified 05/18/2015} is a Portable Document Format (PDF) viewer that is built with HTML5. It is a community-driven project supported by Mozilla Labs. We analyzed version 1.1.1, which has 57,359 LOC, 182 classes, 947 methods, and 876 attributes. Users can interact with the model to access all visualization features and metric values. It is also possible to use drill-down and drill-up operations when one entity is selected. For example, when the user performs a click on the number of classes (\aspas{All classes}) metric, the panel will drill-down to show a list with all the classes. If the user performs a click on one of the classes, the tool will show all information related to the class, i.e., methods, attributes, subclasses, and superclasses. If the user performs a click on \aspas{All methods}, the panel will drill-down again to show a list with all the methods, etc. A menu with all visualization options and diagrams is shown when the user performs a right click on a given element.  

\begin{figure}[ht]
	\centering
	\includegraphics[width=0.9\textwidth]{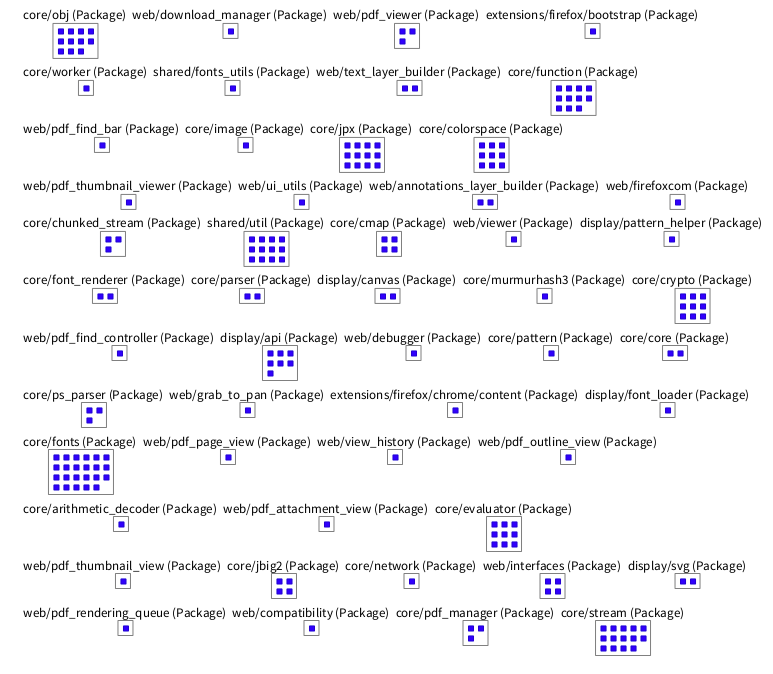}
	\caption{Distribution map for PDF.js (small squares are classes)}
	\label{fig:dm-packages}
\end{figure}

JavaScript does not define language constructs for modules or packages. Therefore, a module can be a single file of JavaScript code that might contain a class definition, a set of related classes, a library of utility functions, or just a script of code to execute~\cite{flanagan2011}. JSClassFinder uses source code files as packages to allow the visualization of distribution maps per packages, like in the example shown in Figure~\ref{fig:dm-packages}. When a user selects the distribution map option, it is possible to choose which parameter will be exposed in the diagram. In this case, packages are represented by external rectangles and the small blue squares are classes. It is also possible to change the colors and to establish a valid range to be considered. For example, users can inform that the diagram should use red squares and consider only classes with more than five methods. This feature can be used, for example, to easily locate the biggest classes in a system. 

\section{Limitations} \label{sec:limitations}

As mentioned before, there are different ways to emulate classes and inheritance in JavaScript. The following specific cases are not covered by our current implementation:

\noindent \textbf{Use of third-party libraries with customized object factories.} For example, ClazzJS\footnote{https://github.com/alexpods/ClazzJS verified on 05/18/2015} is a portable JavaScript library for class-style OOP programming. It provides a DSL for defining classes. Our tool is not able to detect classes nor inheritance relationships implemented with this particular DSL. One idea to make JSClassFinder more robust against this kind of limitation is to gather the most common libraries that provide such service and implement specific strategies to identify them.

\noindent \textbf{Use of singleton objects.} Objects implemented directly, without using any class-like constructor functions, are not considered classes. Listing~\ref{lst_singletons} shows the implementation of one singleton. Although this kind of object is not considered a class, it can be used to compose and clone other objects.

\begin{figure}[htbp]
	\begin{lstlisting}[caption=Singleton object example, label=lst_singletons]
	var person = {	firstName:"John",
					lastName:"Doe",
					birthDate: "01-01-2000",
					getAge: function () { ... }
				}; 
	\end{lstlisting}
\end{figure}

\noindent \textbf{Use of properties bound to variables that are functions.} If a class constructor has a property that receives a variable, it is identified as an attribute, even if this variable holds a function. It occurs because JavaScript is a dynamic and loosely typed language, and JSClassFinder relies on static analysis.

\section{Related Work} \label{sec:related}

There is an increasing interest on JavaScript software engineering research. For example, JSNose is a tool for detecting code smells based on a combination of static and dynamic analysis~\cite{mesbah-scam2013}. One of the code smells detected by JSNose, Refused Bequest, refers to subclasses that use only some of the methods and properties inherited from its parents. Differently, JSClassFinder provides models, visualizations, and metrics about the object-oriented portion of a JavaScript system, including inheritance relationships. Although JSClassFinder is not specifically designed for code smells detection, information provided by our tool can be used for this purpose. 

Clematis~\cite{clematis2014} and FireDetective~\cite{zaidman2013} are tools for understanding event-based interactions, based on dynamic analysis. Their main goal is to reveal the control flow of events during the execution of JavaScript applications, and their interactions, in the form of behavioral models. In contrast, JSClassFinder aims the production of structural models.

ECMAScript definition, in its next version ES6~\cite{ecmascript6}, provides support for class definition. ES6 offers a proper syntax for creating classes and inheritance, similar to the syntax used in some traditional object-oriented languages, such as Java. Since ES6 uses specific keywords in the new syntax, it will be simple to adapt JSClassFinder once the new standard is released. Moreover, JSClassFinder could help on the migration of legacy JavaScript code to the new syntax supported by ECMAScript 6.


\section{Conclusions and Future Work} \label{sec:conclusions}

In this paper we proposed a reengineering tool that supports the identification of class-like structures and the creation of object-oriented models for JavaScript applications. The users do not need to have any prior knowledge about the structure of a system in order to build its model. It is possible to interact with the tool to obtain metric data and visual analysis about the object-oriented portion of JavaScript systems. As future work, we plan to extend JSClassFinder with three major features: (i) support for the new ECMAScript 6 standard, (ii) support to coupling information between classes, (iii) support to the computation of metric thresholds for JavaScript class-like structures~\cite{csmrwcre2014b}. 

\noindent JSClassFinder is publicly available at: \textbf{http://aserg.labsoft.dcc.ufmg.br/jsclasses/}.

\noindent \textbf{Acknowledgment:} This work was supported by FAPEMIG, CAPES and INRIA.

\bibliographystyle{sbc}
\bibliography{thesisbib_CBSoftTools2015}

\end{document}